\documentclass[a4paper]{jpconf}
\usepackage{graphicx}

\begin{document}

\title{Technical aspects in dark matter investigations}

\author{R. Cerulli$^1$,
R. Bernabei$^2$, P. Belli$^2$, F. Cappella$^3$, 
C.J. Dai$^4$, A. d'Angelo$^3$, A. Di Marco$^2$, H.L. He$^4$, 
A. Incicchitti$^3$, X.H. Ma$^4$, F. Montecchia$^{2,5}$,
X.D. Sheng$^4$, R.G. Wang$^4$, Z.P. Ye$^{4,6}$}

\address{$^1$ Laboratori Nazionali del Gran Sasso, I.N.F.N., Assergi, Italy}
\address{$^2$ Dip. Fisica, Univ. di Roma ``Tor Vergata'' and 
INFN, sez. Roma Tor Vergata, Rome, Italy} 
\address{$^3$ Dip. Fisica, Univ. di Roma ``La Sapienza'' and 
INFN, sez. Roma, Rome, Italy }
\address{$^4$ IHEP, Chinese Academy,  P.O. Box 918/3, Beijing, China }
\address{$^5$ Lab. Sperim. Policentrico di Ingegneria Medica, 
Universit\`a di Roma ``Tor Vergata''}
\address{$^6$ University of Jing Gangshan, Jiangxi, China}

\ead{riccardo.cerulli@lngs.infn.it}

\begin{abstract}

Some theoretical and experimental aspects regarding the direct 
dark matter field are mentioned.
In particular some arguments, which play a relevant 
role in the evaluation of model dependent
interpretations of experimental results and in comparisons,
are shortly addressed.

\end{abstract}

\vspace{-0.7cm}

Many Dark Matter (DM) candidate particles 
have been proposed in theories
extending Particle Physics beyond the Standard Model (SM). 
A large number of possibilities is available in different scenarios,
and model frameworks, for particles with very different  
phenomenology and interaction type with ordinary matter. 
A wide literature is available.

From the experimental point of view, many detection processes 
can be considered in order to study the DM particle 
interaction with the target materials.
One of the many cases is the elastic scattering on target nuclei where the 
measured quantity is the nuclear recoil energy.
However many DM candidates can give rise to signals 
that either have totally an electromagnetic nature (see e.g. \cite{damaem}) 
or involve electromagnetic signals associated to 
nuclear recoils (see e.g. \cite{ijma07,chan}); 
obviously, approaches that are based on
multiple subtraction procedure of the electromagnetic component of the
experimental counting rate are blind to these latter scenarios. 
It is worth noting
that also the neutralino in supersymmetric extensions of the SM
could have, in some cases, interaction producing e.m. radiation rather than nuclear recoil.
Moreover, well known side processes
for nuclear recoils exist (such as recoils induced by neutrons, fission fragments, end-range alphas, 
surface electrons, etc.).

In order to interpret the experimental result of a direct search experiment 
many theoretical and experimental parameters and models appear in 
the evaluation and many hypotheses must also be assumed. 
Large possibilities are open
considering the lack of knowledge about the real nature of the candidate, its
distribution in the Galaxy, its coupling to target materials, etc. 
In addition, as regards calculations, one has to consider that:
i)   each model requires its parameters;
ii)  each parameter has an allowed range of values and not just a single value;
iii) uncertainties in the models and in the parameters can play a relevant role
in model dependent interpretations of results and in comparisons.
 
In the following just few of these quantities, models and parameters
-- mostly for the case of DM interactions on nuclei -- are reminded as examples:

\begin{itemize} 

\item  cross section, mass, and other quantities describing the phenomenology and the
       space parameter of the considered DM candidate particle;

\item  Spin-Independent (SI) and/or Spin-Dependent (SD) interaction:
       elastic scatterings  with electromagnetic contribution 
       arising from excitation of bound electrons, inelastic scatterings on target nuclei with either 
       SI and/or SD coupling in various scenarios, 
       interaction of light DM either on electrons or on nuclei with production 
       of a lighter particle, preferred interaction with electrons, conversion of DM 
       particles into e.m. radiation, etc.;

\item  Effective couplings to nucleon: possible isospin dependence of the
       couplings can be considered;  
       effective DM particle-nucleon coupling strengths for either SI
       and/or SD interaction can be defined, and this is important 
       in the comparison of results
       obtained with experiments using different target nuclei depending on the unpaired
       nucleon (compare e.g. odd spin isotopes of Xe, Te, Ge, Si, W
       with $^{23}$Na and $^{127}$I).

\item  Form Factor: it depends on the target nucleus; there is not a universal formulation
       for it, many profiles are available in literature; in these profiles some parameters 
       -- whose value is not fixed -- appear. In case of SD interaction  there is no 
       decoupling between nuclear and DM particles degrees of freedom and  
       it depends on adopted nuclear potential; the form factor profiles can 
       differ one by another by order of magnitude; the value strongly affects the 
       expected signal and the model dependent interpretation of the results;

\item  Spin Factor: in the SD interaction it is a crucial quantity. It depends on the
       nuclear potential;  large differences in the measured counting rate can be expected
       when you consider different target nuclei, different modelizations and when they have
       different unpaired nucleon (compare e.g. odd spin isotopes of Xe, Te, Ge, Si, W
       with $^{23}$Na and $^{127}$I);

\item  Scaling Law: the experimental observable in direct detection when searching for scatterings of DM particles on 
       target nuclei is the nucleus cross section
       of the interaction; in order to compare the results obtained by using different target nuclei
       a scaling law for cross section is required. 
       For example, 
       it has been proposed that two-nucleon currents from pion exchange in the nucleus
       can give different contribution for nuclei with different atomic number \cite{scalaw}; as
       a consequence the cross section for some nuclei can be enhanced with respect to others
       in the SI interactions even within the framework of the MSSM \cite{scalaw}; 
       also this argument has a great relevance in the comparisons;

\item  Halo model and DM velocity distribution: the Dark Halo of the Galaxy is an open problem
       of the field; the expected counting rate for the DM signal 
       depends on the local DM density and on its velocity distribution at Earth™s position. 
       The experimental observations regarding the dark halo of our Galaxy do not allow us to get 
       information on them without introducing models. 
       The uncertainties existing in these models,
       and in their parameters affect the expected counting rate and they must  
       be taken into account. The effect of these uncertainties is very important considering 
       also that each experiment generally use different target 
       material, energy threshold, quenching factor, form factor, etc.. and also the fact that
       the expectation for different DM candidate particle is more sensitive to some of these 
       parameters rather than others. In addition, as regard the velocity distribution,
       the possible existence of non-thermalized component of DM particles in the halo, 
       due e.g. to the tidal stream 
       of satellite galaxies of Milky Way, can give relevant contribution and 
       impact in results and comparisons.
      
\end {itemize}

For some of the many other complementary aspects regarding uncertainties in the experimental results
and in the comparison see e.g. ref. \cite{taupnoz}.

The DM direct experiments can be classified in two classes:
experiments investigating a peculiar model independent DM signature able to point 
out the presence of a DM contribution in their measured rate, and experiments
based on the comparison of their rate or of their recoil-like candidates surviving many subtractions
with an expectation calculated in one assumed scenario (which implies to adopt
many assumptions and approximations). 

As regard the first class of experiments, at present the only feasible model
independent signature is the DM annual modulation exploited by the DAMA experiments at the 
Gran Sasso National Laboratory \cite{RNC,modlibra,modlibra2}. This experiment
with the present DAMA/LIBRA and the former DAMA/NaI 
(exposed masses: $\simeq$ 250 kg and $\simeq$ 100 kg of highly radiopure NaI(Tl), respectively) 
set-ups has released so far a total exposure of 1.17 ton $\times$ yr
over 13 annual cycles, obtaining a positive model
independent evidence for the presence of DM particles in the galactic 
halo at 8.9 $\sigma$ C.L.. No systematics or side reactions 
able to mimic the signature (that is, able to
account for the measured modulation amplitude and simultaneously satisfy
all the requirements of the signature) has been 
found or suggested \cite{RNC,perflibra,modlibra,modlibra2,add}.
In the last years CoGeNT experiment
with a P-type Point Contact Germanium detector reported a preliminary 
positive hint \cite{CoGeNT}. 

As regard the second class, these experiments, 
in order to reduce their experimental counting rate, generally
perform huge data selections and many subtraction procedures;
after all, they derive a set of recoil-like candidates.
It is worth noting that not only uncertainties in the many applied 
large selection/subtraction procedures 
and in the related efficiencies are present, but well known side reactions exist giving 
recoil-like candidates surviving all the applied subtractions.

The use of different techniques and 
experimental approaches,
different target materials and 
experimental conditions must be considered in the comparisons. 
Some general requirements for DM experiments must be satisfied. 
In particular, it is mandatory for an experiment to accurately
know the energy scale and energy resolution in the keV energy range of interest;
this is crucial especially for dis-uniform detectors (as the liquid noble gas ones;
see their and other literature).
In addition, it is mandatory
to know the response of the detector at threshold validated by careful 
calibrations performed in the same condition as in the production runs. 
In some experiments, as e.g. DAMA/LIBRA, the energy scale and resolution  are continuously 
measured by external/internal known 
sources from MeV down to the energy threshold \cite{perflibra}
while in other experiments these quantities, in the low energy region of interest
for DM investigation, are extrapolated from calibrations at much higher energy 
(see e.g. \cite{paperliq}). 

Another important quantity -- in the case of elastic scattering on target nuclei --
in results and comparisons
is the quenching factor, $q_f$ and, in particular for double phase experiments, 
the L$_{Eff}$ parameter, and their behavior at low energy. 
As regard L$_{Eff}$, tensions exist in the available measurements and also
in theoretical estimation (see e.g. \cite{collar}). The sensitivity for DM particle of double phase experiments 
strongly depends on this quantity and a cautious approach in its extrapolation
is mandatory in order to not overestimate experimental sensitivity. 
The quenching factor is a specific
property of the employed detector(s) and not a general
quantity universal for a given material. For example, in
liquid noble-gas detectors, it depends, among other things,
on the level of trace contaminants which can vary in time
and from one liquefaction process to another, on the
cryogenic microscopic conditions, etc..
In bolometers the quenching factor 
depends for instance on specific properties, trace contaminants,
cryogenic conditions, etc. of each specific detector,
while generally it is assumed exactly equal to unity. 
In scintillators, the quenching factor depends, for example,
on the dopant concentration, on the growing method/
procedures, on residual trace contaminants, etc., and is
expected to have some energy dependence. 
The uncertainty of $q_f$ value is not easy to be deeply evaluated in neutron  measurement
and it can be large. In the last years it has been proposed a general empirical formula to derive 
the quenching factor values for each used detector \cite{tret}. For example   
it derives quenching factors values for NaI(Tl) larger at low energy than 
those usually adopted; in addition, the possible effect of channeling has also to be 
considered \cite{chan}. These two effects can play an important role in the interpretation of 
results and in comparisons.

As regard problems related to the application of multiple subtraction procedures 
of the measured counting rate, as pursued by experiments trying to identify 
the presence of recoil-like events,
some cases will be mentioned.

In particular, for the rate claimed by XENON-100 in an inner "fiducial" volume 
some arguments can be pointed out: 
i) the energy calibration of the underground detector has been carried out in significantly
higher energy region than the claimed keV one, and
thus the energy threshold, the energy scale, the energy resolution and all the necessary
efficiencies for the many applied subtraction procedures are unproved;
ii) the small number of available photoelectrons/keV in the underground detector, the huge
detector disuniformity, the far UV scintillation light emission, etc.
do not support that energy threshold;
iii) the stability levels of all the cuts windows applied in the many subtraction procedures
and of the related efficiencies have not been suitably proved; 
iv) the reproducibility of identical detector performances
in different Xe liquefaction
processes in the relatively large underground volume (as well as along production runs) and 
the needed level of  stability of all
the detector features during each data taking period have not been suitably proved as well;
etc.
Some other general aspects have been addressed in \cite{paperliq,taupnoz,collar}.

As regards recent results by CDMS-II \cite{cdms}, a huge selection of the available
detectors and data is performed and many subtraction procedures are applied as well, further reducing to
marginal the already very limited available exposure; the stabilities of all the applied cuts windows
are unproved, the quenching factor is always assumed equal to unity, all the efficiencies
values and their stability are not suitably proved as well, etc.

Finally it is worth noting that the limits achieved by experiments
searching for DM particle giving rise to nuclear recoils, 
are not robust reference points. In fact
as discussed before similar results are quite uncertain
not only because of possible underestimated or unknown systematics in the huge data subtractions
and in some experimental aspects, but also because the results refer only to 
a certain (generally largely arbitrary) set of assumptions.
When one consider, for example, the energy threshold dependence of the 
exclusion plots, up to several orders of
magnitude differences can be present between claimed and realistic
evaluation of the experimental sensitivity. 

In conclusion, all the arguments mentioned above allow one
to understand that the uncertainties existing in the
models and in their parameters, the sometimes not suitably determined experimental quantities,  
etc., can drastically affect model dependent results and comparisons.
Thus the comparison of the results achieved by different experiments
must be handled with very cautious actitude without neglecting the many sources of 
uncertainties. This holds in particular for model dependent exclusion plot, that cannot 
be considered as universal limit. 
Model independent approaches are mandatory.

\end{document}